\documentclass[12pt]{revtex4}
\usepackage[latin1]{inputenc}
\begin{document}
\author{Alejandro Rivero}
\affiliation{Institute for Biocomputation and Physics of Complex Systems, University of Zaragoza, Spain}
\email{al.rivero@gmail.com}
\title{Unbroken supersymmetry, without new particles}
\begin{abstract}
We consider Koide's equation for charged leptons jointly with 
the hypotheses implied in the ancient Dual Quark-Gluon model. The focus
is the possibility of motivating a supersymmetric framework
akin to the one of D=11 maximal supergravity.
\end{abstract}
\maketitle

The dual Quark-Gluon Model of Hadrons, where mesons and quarks are in equal footing, was 
proposed early \cite{schw1,schw2} as a consequence of Ramond model of fermions\cite{ramond}.
This model was obvious if one is willing to encompass all composite QCD strings (mesons and diquarks)
and fermions (baryions) in a superstring model, as the emission of a diquark barionic
string from a baryion will produce a single quark state.

It was forgotten after the discovery of supergravity and the reinterpretation
of dual models as a theory of gravity. Also, the theoretical dificulties of working
with a model both elementary and composite of itself was evident, and it was advocated
to see it as "a quark model without quarks", neglecting the fermion side. As it has
been told elsewhere\cite{arivero}, the model only closes on itself for three
generations and one highly massive top quark, or for more than three generations with
an absurd number of massive tops and bottoms. Back in 1971, there was no experimental
support for a third generation, so this prediction (never written down nor
proposed, as far as I know) was a fail of the model.

\section{Koide's model of subquarks.}
In 1981, Y. Koide \cite{koide} suggested some models where quarks and leptons were composed of more elementary bosons and fermions having global $SU(3)$ color and $SU(3)$ ''generations'' symmetry. With an adequate choosing of the representations for these subparticles, plus a relatively adhoc symmetry breaking scheme, it was possible to derive some parameters of the CKM matrix and, more interestingly, to predict the mass of the tau lepton which was measured years later. The prediction has today an astonishing precision within the current experimental error.

Koide's symmetry breaking scheme constrains the square root of the mass matrix $M_l$ of charged leptons. When $M^{1/2}_l$ is decomposed in a central part $U$ (a multiple of the identity) plus a traceless part $V$,\begin{equation}
(U+V)^2=M_l
\end{equation}
the symmetry breaking scheme imposes a relationship between traces
\begin{equation}
Tr[U^2]-Tr[V^2]=0
\end{equation}
 And it works: with PDG 2009 data, the LHS is between -0.05 and 0.09 MeV. Or, if you prefer an adimensional quantity,
the quotient $Tr[U^2]/Tr[V^2]$ is $1.00002 \pm 0.00008$.

Recently, Koide has produced some ways to the same formula without asking for compositeness, 
but we keep an eye on it because of the next section. It must be stressed that in 1981 the
value of the mass of the tau lepton was far away from its current measurement.

I want to do two observations here:

1) If we think of mass as a component of the momentum operator $P_\mu$, the fact of having a condition in its square root smells to a condition on supersymmetry generators.

2) If supersymmetry were unbroken, the same condition should appear in the scalar partners of the charged fermions. Furthermore, we could think that the sfermions are composite. Even we could think that the sfermions are the only composites, where Koide's breaking scheme applies, and that Koide's formula in the leptons is only a reflection of the actual formula for sfermions.

\section{sBootstrap.}

A motivation to pursue the above idea is that the masses of muon and tau, which are free parameters in the standard model, are very near of similarly charged (pseudo)scalar particles, the pion and the D meson. In fact the former case is so near than historically the muon was first interpreted, mistakenly, as a pion. 

We advance one step over Koide's models of subquarks and do a bold suggestion: that the particles labeling the sfermions are really quarks themselves.

Astonishingly we exceed expectations when we notice that there are actually five "light" quarks, on the QCD-Chiral-Electromagnetic mass scales, and a massive quark, in the electroweak mass scale. And this quark is as massive as to be unable to bind into mesons. So the flavour global symmetry of our Standard Model quarks is SU(5). It can  be labeled under SU(3)xSU(2) in order to separate the (d,s,b) and (u,c) kinds of charge.

The decomposition of SU(5) is well known. Take the 24 of $5 \otimes \bar 5 = 24 \oplus 1$ and the $15$ of $\bar 5 \otimes \bar 5 = 15 \oplus 10$. Then
under  $SU(5) \supset SU(2) \times SU(3) \times U(1)$
we have
\begin{eqnarray}
15&=&(3,1)(6)+(2,3)(1)+(1,6)(-4) \\
24&=&(1,1)(0)+(3,1)(0)+(2,3)(-5)+(2, \bar 3)(5)+(1,8)(0)
\end{eqnarray}

The $(2,3)(1)$ and $(1,6)(-4)$ are the partners of quarks down and up respectively. The antiparticles are provided in the $\bar {15}$ representation.

The $(2,3)(-5)$ and $(2, \bar 3)(5)$ are the partners of positron and electron. The other 12 particles of this multiplet are neutral.

So our \cite{arivero} basic observation here is  that

3) For three generations and a single "massive" quark, the system closes on itself: the degrees of freedom generated in the product are the ones needed for the sfermions of a supersymmetric standard model, which in turn are transformed by susy into the original fermions we need to generate them.

It is not possible to do the same trick with any other number of generations, so in some sense the sBootstrap fixes the number of generations.

A major objection against the sBootstrap mechanism is that, having paired the mesons from QCD with
the elementary quarks and leptons, we should expect to detect an string structure in the
electron spin and charge, which we do no detect. On the other hand, it is more satisfactory to think
that there is some principle forbidding fermions to have extended non-pointset structure, that
to think that string theory has provided us with a local extended model of the electron.
It could be worth to remember (Susskind et al.) 
that the initial research in dual models run into problems when calculating the sizes 
of hadronic strings.

\pagebreak
\section{D=11 supergravity multiplet.}

There are some general motivations to look for susy in D=11 instead of down-to-earth in D=4. First,
it is well known that the minumum number of extra dimensions to allow for SU(3)xSU(2)xU(1) symmetry
is 7. It is known that superstring theory lives in D=10 and develops an extra dimension in some
limits. It is less known that Connes' geometric version of the standard model lives in dimension
2 mod 8. 

Superstring theory is a motivation for us from the point of view of our composites: we really would
like to interpret each of our sfermions as quarks at the ends of a string. Really we know from
Sagnotti and Marcus that the SO(32) symmetry of the quantised superstring is produced with only
a set of 5 ``fermion flavour'' in the worldsheet.

Also there is an intriguing ``reggeization'' of the process of $Z^0$ decay, which happens
to show a total rate coinciding with the scaling of electromagnetic decay of QCD strings (refer 
to the dimensionless quantity $\Gamma^{e.m.}_X/m^3_X$, for X a QCD meson).

Supergravity is a motivation because its minimal multiplet has barely the number of degrees
of freedom to store the information of the supersymmetric standard model, except for the Higgs.
See \cite{boya}: after accounting for all the standard model, their gauge forces and their 
superparners, plus the 2 degrees of freedom of the 4-D gravitino, only 6
scalars, from the 128 of the full multiplet, are left. And we need 8 for the Higgs of the
MSSM. Yet, the Higgs mechanism is still undiscovered. Or, the graviton could be exiled in
favour of a 4D, $N=4 \times N=4$, superYangMils theory.

D=11 supergravity has a natural way to produce D=4 because of the structure of supersymmetry
multiplets. Its 128-component fermion can not be matched with a 44 component graviton only,
so an extra field is needed to hold the extant 84 bosons. This field happens to be a 
tensor with three indices and thus induces a 4D uncompactified sector in
some models.  

Amusingly, we could be interested on two numerological happenings of the number "84" in the standard model: 

- the "charged fermions" of the standard model amount to 84 degrees of freedom (plus 12 neutrinos = 96).
They could be organised in three families of multiplets of 28 components each.

- the "light fermions" of the standard model amount to 84 degrees of freedom (plus 3 colors of the top quark = 96). They could be organised in two chiral multiplets of 42 components each, or four multiplets of 21 components.

The first realisation seems easier to look for. Barring the neutrinos, we expect a non chiral 
theory. Strings of type IIA, which are known to recover 11D supergravity in the strong coupling
regime, have a 28 in the NS-NS sector and a 56 in the R-R sector, coming both from the 
decomposition of the 11D 3-form.

If we refine down to 9D via compactification in $S^1$, and even having the fact that 9D is 
non-chiral, we have some scent of the organisation on 21-plets: the 28 decomposes in a 21 + 7, and
the 56 in 21+35. Of course we want the 9D theory to be non chiral: it has the minimum number of extra dimensions required to fit in Kaluza Klein a SU(3)xU(1) theory, which is the non-chiral part of the standard model. 

In the standard model, the mass of the top quark seems linked to the EW breaking scale, having
an experimental yukawa coupling equal to one, within a few percent. So if we take the EW scale
to infinity, completely breaking the group, then also the top is removed from the scenary. This
provides some motivation to look for the second realisation.

A more than bold conjecture should be that this second realisation hides in IIB. Being chiral,
IIB theory does not descent from 11D SUGRA. The 84 degrees of freedom are hidden in the sum
of a 28 in the NS-NS sector and a 28 plus a 35 in the R-R sector, ie two 2-forms and one
self-dual 4-form. To produce them explicitly, it is needed to consider the 9D theory 
and to discard a mix of the 9th component of both 2-forms, accounting for 7 components. 
In this way we are left with a 35, two 21s, and the surviving 7 components of the mix. Of course,
noticing that IIA amounts to include the 7 components from the R-R sector, in this case
we could be tempted to unmix and directly exclude them, taking the ones from the NS-NS sector.


What is left to find is a way to produce the SU(2) group in string theory. It could be time to
revisit the symmetry enhancement condition which happens when the compactifyied radius
is equal to the string scale. Fixing $R=\alpha'^{1/2}$ and moving R to zero needs a careful
consideration of the limit process, because at the same time we enter into the
strong coupling regime, where IIA grows an extra dimension.

\section{Koide's relationship for mesons}

We can think that the supersymmetry in the bosonic part has got an extra contribution from the 
breaking of flavour symmetry. So lets see what happens if we restore some of this 
breaking.

If we set the masses of the bottom and strange quark to the same value that the mass of the
down quark, we are left with only three mass levels: 1870 MeV, 139.5 MeV and 0. The zero
level should be inhabited by the ``$\eta_1$ singlet'' combinations of $\pi^-, K^-, B^-$ 
and $D^-, D^-_s , B^-_c $, while the other levels are 
inhabited by the ``$\eta_8, \pi_8$" combinations.

SUSY, of course, is still slightly broken, if we compare with the lepton
triple (0.511, 105.66, 1777 MeV). But Koide's relationship fares better, and the
quotient $Tr[U^2]/Tr[V^2]$ is about 1.005. It seems that the breaking
of SUSY aims to preserve Koide's, contrary to our initial expectation (or,
again, exceeding it)

If instead restoring SU(3)-down flavour we restore SU(2)-up flavour (or none at all) we still have 
that the masses of Kaon and Upsilon, jointly with a zero mass, also form a Koide ``triple".  
In fact there are no more triples with a zero mass. 

In the neutral case the fit is not so good because the Kaon is a highly mixed particle. 
The $\eta_8$ of the classical SU(3) ``u,d,s" flavour group is a better candidate.

Other triples in the meson and baryon sectors have been explored by Carl Brannen \cite{brannen}. 

\section{Puzzling chirality}

Of course, there are no chiral fermions in D=11, thus no way to implement SU(2) in the way it works in the standard model. It seems that the chiral part of the standard model, $SU(2) \times U(1)_{ew}$, interpolates, as we move the mass of W from zero to infinity, between two 
non-chiral theories in D=9 and D=11. 

The neutrinos in the 24 of SU(5) appear in a very irregular way: a triplet, a singlet and 
an octuplet. We need to mix them if we want to recover a grouping in three generations.
From the point of view of composites $D\bar U$, the charged leptons can contemplate two
ways to decay to neutrinos: either a decay $D$ to $U$, landing in the triplet, or a decay
$\bar U$ to $\bar D$, landing in the combination of octet and singlet. So perhaps
the (3,1) triplet of the $24$ has some special role to build chiral interactions.

The (3,1) triplet from the $15$, and its antitriplet from the $\bar 15$, are the only predicted particles not in the standard model. They should have electric charge 4/3 so it seems that we can not accomodate them as the partners of (three generations of) a two-component fermion. But we can not arrange them in Dirac fermions, as they amount to 6 degrees of freedom, or 18 if we consider that they should be coloured as the rest of the $15$ multiplet. More, if they are coloured they overcrown the D=11 multiplet. So it seems that the (3,1) triplet in $15$ asks for a chiral fermion in each family, but the sBootstrap, charge, and other considerations, ask for this chiral fermion to dissapear.

\end{document}